\documentclass[RNAAS]{aastex62}

\newcommand\ktwo{\textit{K2}}
\newcommand\tess{\textit{TESS}}
\newcommand\kepler{\textit{Kepler}}

\graphicspath{{./}{figures/}}

\begin{document}

\title{A catalog of stars observed simultaneously by Kepler and \tess}


\author[0000-0001-7139-2724]{Thomas Barclay}
\affiliation{NASA Goddard Space Flight Center, 8800 Greenbelt Rd, Greenbelt, MD 20771}
\affiliation{University of Maryland, Baltimore County, 1000 Hilltop Circle, Baltimore, MD 21250, USA}

\author[0000-0002-3306-3484]{Geert Barentsen}
\affiliation{NASA Ames Research Center, Moffett Field, CA 94035}
\affiliation{Bay Area Environmental Research Institute, P.O. Box 25, Moffett Field, CA 94035, USA}

\keywords{surveys, catalogs}

\section{}
The \kepler\ spacecraft provided the first long-baseline, high-precision photometry for large numbers of stars \citep{Borucki2010}. This enabled the discovery of thousands of new exoplanets \citep[e.g.][]{Rowe2014,Morton2016,Thompson2018,Mayo2018}, and the characterization of myriad astrophysical phenomena \citep[e.g.][]{Bedding2011,Thompson2012,Garnavich2016}. However, one of the challenges with interpreting \kepler\ data has been that no instrument has provided a comparison dataset. Therefore, the replication of \kepler\ time-series data has remained elusive. While there have been efforts to reproduce observations of \kepler's transiting planets using the \textit{Hubble} and \textit{Spitzer Space Telescopes}, this has been limited to a small number of stars and short observing baselines \citep{Beichman2016}.

The \textit{Transiting Exoplanet Survey Satellite} (\tess) launched in April 2018 and began science operations in July 2018 \citep{Ricker2015}. During the \tess\ primary mission, it will survey 85\% of the sky. NASA's first two observatories dedicated to discovering exoplanets, \kepler\ and \tess, were simultaneously operating during 2018. While the \ktwo\ mission surveyed the ecliptic plane, \tess\ targets fields outside the ecliptic. However, during September 2018, a small region of the sky was observed simultaneously by both \tess\ and \kepler\ as part of \tess's Sector 2 (Aug 22 - Sep 20, 2018) and \kepler's \ktwo\ Campaign 19 (Sep 7 - Sep 26, 2018). The overlap region was 0.5~deg$^2$.

\tess\ full-frame image mode collected the entire region, but \ktwo\ targets must be pre-selected. We used the Web TESS Target Tool \citep{Mukai2017}, available at the TESS Science Support Center website, to determine which \ktwo\ Campaign 19 targets\footnote{The K2 target lists are available at \url{https://keplerscience.arc.nasa.gov}.} were observed by \tess.
We identified 171 \ktwo\ targets that fell inside \tess\ Sector 2 field of view, all of which are observed at 30-minute cadence by both missions. We cross-matched these 171 targets to the \tess\ Input Catalog \citep{TIC} and found 169 matches within 6''. The two sources with no match were fainter than Kp=18. A portion of the catalog is shown in Table~\ref{tab:1}.

The targets range in brightness in the \tess\ bandpass ($Tmag$) from 6.7--18.4 ($Kp$=7.7--19.2), with 93 targets brighter than $Tmag$=15, and 17 brighter than $Tmag$=12. The majority of the sample are main-sequence stars. We cross-matched the sample with Gaia and found 153 matches \citep{Gaia2016,Gaia2018}. Using the Gaia distance estimates from \citet{Bailer2018}, we found targets range in distance from 80--6500 pc, with a mean distance of 800~pc. The brightest target is the multiple star system HD 218928 ($Kp$=7.7). The sample also contains approximately 30 galaxies which were observed for the \ktwo\ Supernova Experiment. There is a Seyfert galaxy (EPIC 245934248), a source with high-proper motion (EPIC 245924999, 180 milliarcsec/yr), a white dwarf (EPIC 251456407), and an RR Lyra star candidate (EPIC 251813843). Of these 171 targets, 81 were previously observed during \ktwo's Campaign 12 (Dec 2016 - March 2017).

This dataset provides the first simultaneous, long-duration, high-precision observations of the same targets from different space-based observatories and presents an excellent opportunity to explore instrumental systematics present in the two telescopes. For \kepler, we can look to see the extent to which the optics, detectors, or electronics are contributing signals to the data, and disentangle these from astrophysical contributions. For \tess\ we can use \kepler\ data, with its considerably higher photometric precision, as a ground-truth. 

\begin{deluxetable}{lllllllllll}
\tablecaption{A sample of the targets observed simultaneously by \kepler\ and \tess\label{tab:1}}
\tablehead{
\colhead{EPIC ID} & \colhead{TICID} & \colhead{\textit{Gaia} DR2 ID} & \colhead{RA} & \colhead{Dec} & \colhead{$Kp$} & \colhead{$Tmag$} & \colhead{$G$} & \colhead{Teff} & \colhead{Radius} & \colhead{Dist}\\
 &  &  &  &  &  &  &  & \colhead{(K)} & \colhead{($R_\odot$)} & \colhead{(pc)}
}
\startdata
245925582 & 4554210 & 2413051302797084416 & 347.998206 & -11.933554& 7.66 & 6.68 & 7.49 & 5047 & 9.55 & 253\\
245929348 & 4579916 & 2413014469157323264 & 348.407601 & -11.749342 & 9.32 & 8.10 & 9.02 & 4130 & 23.55 & 702\\
245931711 & 4582469 & 2413392224416454528 & 348.516775 & -11.629405 & 9.94 & 9.41 & 9.79  & 6296 & 1.18 & 148\\
245927251 & 4520668 & 2413067451874115840 & 347.867863 & -11.852871 & 10.93 & 10.51 & 10.83 & 6736 & 1.85 & 429\\
\enddata
\tablecomments{Table 1 is published in its entirety as a comma-separated values file. A portion is shown here for guidance regarding its form and content. $Tmag$ is from the \tess\ Input Catalog \citep{TIC}, RA, Dec and $Kp$ are from the \ktwo\ Ecliptic Plane Input Catalog \citep{Huber2016}, G, Teff and radius are from {\it Gaia} DR2 \citep{Gaia2016,Gaia2018}, and distance is inferred from {\it Gaia} DR2 \citep{Bailer2018}. This catalog is also available from Figshare \citep{catalog}.}
\end{deluxetable}  

\begin{figure}
    \centering
    \plotone{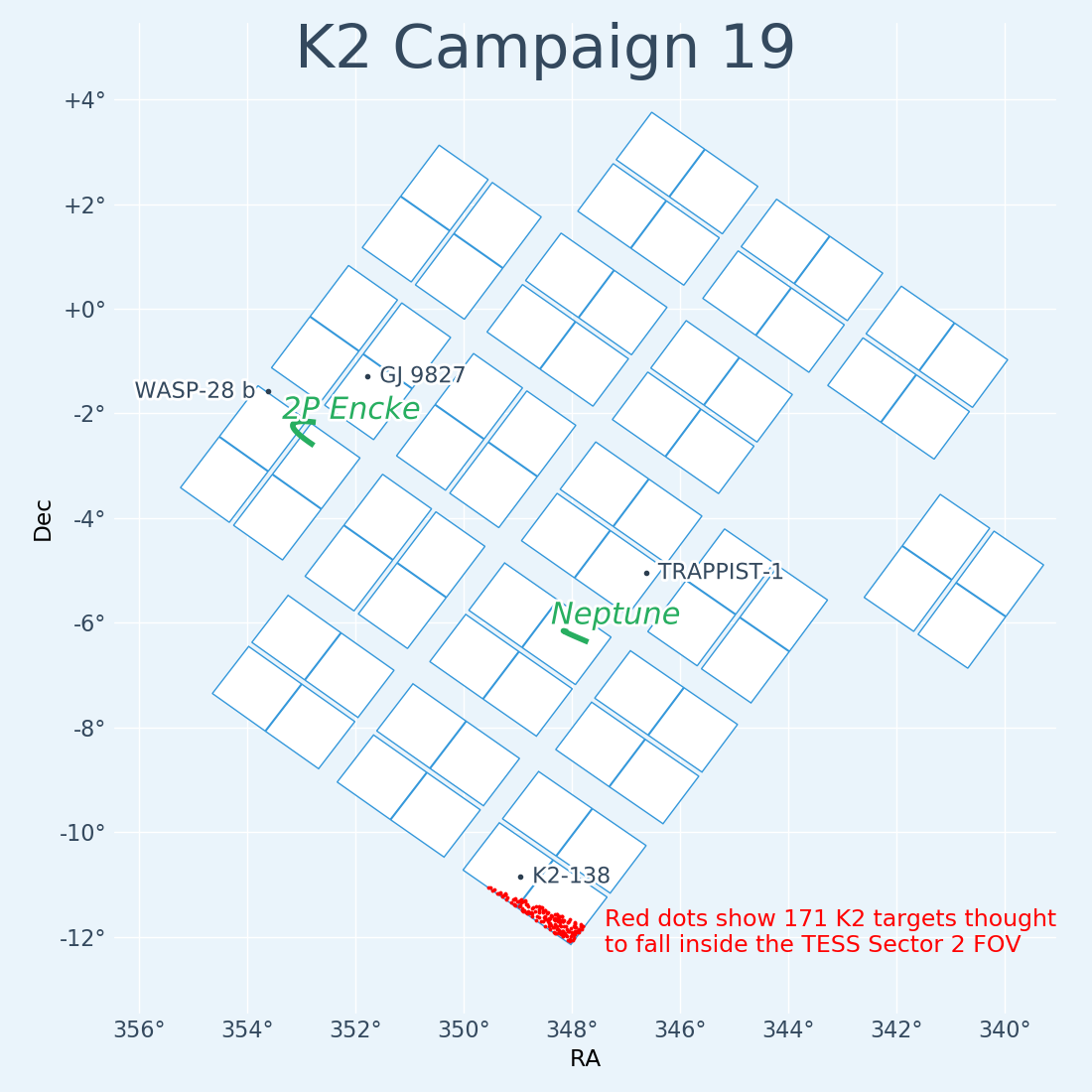}
    \caption{The location of \ktwo\ Campaign 19. In red are the 171 targets that overlap with the field of view of \tess\ Sector 2. This figure is not included in the RNAAS published version of this paper.}
    \label{fig:1}
\end{figure}

\acknowledgments
We thank the PIs of \ktwo\ programs with targets in our catalog: Matthew Burleigh, William Cochran, Courtney Dressing, Ryan Foley, Joyce Guzik, J. J. Hermes, Andrew Howard, Daniel Huber, Adam Jensen, Marshall Johnson, Armin Rest, Robert Szabo, Dennis Stello, and Thomas Barclay. This work has made use of data from the European Space Agency (ESA) mission {\it Gaia} (\url{https://www.cosmos.esa.int/gaia}), processed by the {\it Gaia}
Data Processing and Analysis Consortium (DPAC, \url{https://www.cosmos.esa.int/web/gaia/dpac/consortium}). Funding for the DPAC has been provided by national institutions, in particular the institutions participating in the {\it Gaia} Multilateral Agreement.

\software{
IPython \citep{ipython},
Jupyter \citep{jupyer},
Pandas \citep{pandas},
Astropy \citep{astropy},
Astroquery \citep{astroquery},
TopCAT \citep{Topcat},
K2fov \citep{k2fov},
tvguide \citep{Mukai2017}
          }

\bibliography{tess}

\end{document}